\documentclass[prd,preprint,a4paper,superscriptaddress,nofootinbib,11pt]{revtex4}
\usepackage{amsmath,amsfonts,amssymb}
\usepackage{txfonts}
\usepackage{epsfig}
\usepackage{graphics}
\usepackage[T1]{fontenc}
\usepackage[safe]{textcomp}
\usepackage{dsfont}

\usepackage{csquotes}

\begin{document}

\begin{center}

{\bf \Large Dimensional flow in the kappa-deformed space-time}
 
\bigskip

\bigskip

Anjana V. {\footnote{e-mail: anjanaganga@gmail.com}}and E. Harikumar  {\footnote{e-mail: harisp@uohyd.ernet.in }} \\
School of Physics, University of Hyderabad,
Central University P O, Hyderabad-500046,
India \\[3mm] 

\end{center}

\setcounter{page}{1}
\bigskip

\begin{center}
 {\bf Abstract} 
\end{center} 
We derive the modified diffusion equations defined on kappa-space-time and using these, investigate the change in the spectral dimension of kappa-space-time with the probe scale. These deformed diffusion equations are derived by applying Wick's rotation to the $\kappa$-deformed Schr\"odinger equations obtained from different choices of Klein-Gordon equations in the $\kappa$-deformed space-time. Using the solutions of these equations, obtained by perturbative method, we calculate the spectral dimension for different choices of the generalized Laplacian and analyse the dimensional flow in the $\kappa$-space-time. In the limit of commutative space-time, we recover the well known equality of spectral dimension and topological dimension. We show that the higher derivative term in the deformed diffusion equations make the spectral dimension unbounded (from below) at high energies. We show that the finite mass of the probe results in the spectral dimension to become infinitely negative at low energies also. In all the cases, we have analysed the effect of finite size of the probe on the spectral dimension.

\newpage
 
\section{Introduction}
Combining the principles of quantum mechanics and general relativity is known to result in space-time uncertainties\cite{dop}. This leads to fuzziness of the space-time at extremely short distances. This fuzziness can change the effective dimension of the space-time at high energies\cite{dimred1, dimred2, dimred3}. Various approaches like string theory, loop gravity and causal dynamical triangulation have been developed to unravel the nature of space-time at extremely short distances. All these approaches have a common trait that they predict dimensional reduction\cite{string, loop, dm, cdt, satheesh}. Construction and analysis of diffusion equation, compatible with these approaches is a possible way to study the dimensional flow. Spectral dimension turn out to be an important tool for investigating the nature of space-time at microscopic scales\cite{dm}.

Non-commutative geometry is a possible way to capture the space-time uncertainties and thus study the space-time structure at Planck scale. $\kappa$-space-time is an example of a Lie algebraic type non-commutative space-time whose coordinates satisfy
\begin{equation}
[\hat{x}_0, \hat{x}_i]=ia\hat{x}_i,~~~~~ [\hat{x}_i, \hat{x}_j]=0 .
\end{equation}
Significance of this space-time to quantum gravity comes from the fact that it appear naturally in the low energy limit of loop gravity\cite{quantum} as well as in the context of doubly-special relativity theories\cite{ds}. Analysis of diffusion on this space shows that the effective dimension is different from the topological dimension\cite{nc1,nc3,nc2,amelino,sdkappa}.

In this paper, we investigate the dimensional flow in the $\kappa$-deformed space-time using the solution of deformed diffusion equations. These $\kappa$-diffusion equations are constructed by a Wick's rotation of $\kappa$-deformed Schr\"odinger equations obtained as the non-relativistic limit of the well studied $\kappa$-deformed Klein-Gordon equations\cite{sm, 17a, 17b, kgeqn, twist}.

It is well known that Schr\"odinger equation and diffusion equation are related by a Wick's rotation\cite{book1}. The time-dependent Schr\"odinger equation for a free particle is 
 \begin{equation} 
   i\hbar\frac{\partial}{\partial t}\phi(x,t) = -\frac{\hbar^2}{2 \mu} \nabla^2 \phi(x,t) \label{schroeqn}
 \end{equation}
where $\mu$ is the particle's reduced mass, $\nabla^2$ is the Laplacian, and $\phi$ is the wave function, is mapped to  
 \begin{equation}
   \frac{\partial}{\partial t}\phi = k \nabla^2 \phi,
 \end{equation} 
under the map $t \rightarrow -it$. By re-defining $kt=\sigma$, one re-express the above equation as the standard diffusion equation .  

The analysis of the effective dimension of the space-time and its dependence on the probe scale is studied using a diffusion process\cite{dm}. In this approach, one investigate the behaviour of a non-relativistic particle undergoing diffusion in the space whose dimension is under study. This equation is solved by imposing the delta function initial condition which takes into account of the point particle nature of the probe.

The motion of the non-relativistic particle in a diffusion process in d-dimensional space-time is governed by the diffusion equation
 \begin{equation}
   \frac{\partial }{\partial \sigma}U(x,y;\sigma)  =  \mathcal{L}U(x,y;\sigma)\label{de}
 \end{equation}
where $\sigma$ is the diffusion time, $ \mathcal{L} $ is the generalized Laplacian in the given space (of d-1 dimensions) and its solution U(x,y; $\sigma$) is the probability density of diffusion from x to y during the diffusion time $\sigma$. Using the solution of the diffusion equation, one finds the return probability as
 \begin{equation}
   P(\sigma)= \frac{\int d^n x \sqrt{det g_{\mu \nu}}U(x,x;\sigma)}{\int d^n x \sqrt{det g_{\mu \nu}}}. \label{p_g}
 \end{equation}
 The logarithmic derivative of return probability $P_g$($\sigma$) gives us the spectral dimension of the corresponding d-1 dimensional space, i.e., 
 \begin{equation}
   D_s = -2 \frac{\partial \ln P(\sigma)}{\partial \ln \sigma}.\label{d_s}
 \end{equation} 

In the approach of studying the dimensional flow of space-time\cite{dm,nc1,nc3,nc2,sdkappa,amelino}, one uses the non-relativistic diffusion equation given in eqn.(\ref{de}) but replaces the d-1 dimensional Laplacian $ \mathcal{L} $ with the Euclideanised Beltrami-Laplace operator defined on the concerned space-time. One also interpret `$\sigma$' as the fictitious diffusion time. The quantum gravity effects do modify the Beltrami-Laplace operator, which is typically the kinetic part of the deformed field theory, defined on the space-time under study. An equivalent approach, using the momentum space representation of the kinetic part of the deformed field theory (which is essentially the deformed energy-momentum relation) has also been used to study the spectral dimension of various models\cite{visser}.

There have been attempts to study possible generalizations of the diffusion process described by the eqn.(\ref{de}), which includes changes in the Beltrami-Laplace operator, modification in the initial conditions as well as the modification of diffusion operator $\frac{\partial}{\partial \sigma}$\cite{prob}, in order to capture possible quantum gravity effects. Modification of diffusion equation to address the non-trivial scaling behaviour of space-time was analysed and it was also shown that the diffusion equation do get modified by introducing diffusion in non-linear time as well as by incorporating a non-trivial source term\cite{prob}. Natural generalization of diffusion equation involving fractional derivatives ( in spatial coordinate as well as in diffusion time ) was also introduced and discussed\cite{prob}.  

The spectral dimension of $\kappa$-Minkowski was studied in\cite{nc1} and the fractal nature of space-time with quantum group symmetry was exhibited for the case of Wick's rotated $\kappa$-Minkowski space. Casimir of $\kappa$-Poincare algebra was used to calculate the trace of the heat kernel in Wick's rotated $\kappa$-Minkowski space. The numerical evaluation of resultant expression showed that the spectral dimension change from 4 to 3 with the probe scale. A study on similar lines is reported in\cite{nc3}. Here, the spectral dimension was studied using three possible forms of the $\kappa$-deformed Laplacians in the momentum space. The Laplacians conceived from the casimir of the bi-covariant differential calculus displayed the dimensional reduction as the time change from 4 at low energies to 3 at high energies. For the Laplacian associated with bi-crossproduct Casimir, spectral dimension varies from 4 to 6 with energy. A model compatible with the notion of relative locality\cite{relo1, relo2, relo3, relo4} gives spectral dimension that goes to infinity as one move to UV regime\cite{nc3}.

The spectral dimension of $\kappa$-space-time using the $\kappa$-deformed diffusion equation was studied in our earlier work\cite{sdkappa}. Using a mapping of non-commutative coordinates to commutative coordinates and their derivatives, we constructed the diffusion equation in $\kappa$-space-time from the Casimir of the undeformed $\kappa$-Poincare algebra\cite{sm}. Keeping terms upto second order in the deformation parameter $a$, we solved the diffusion equation perturbatively. The spectral dimension derived from this solution showed a length scale dependence. For a 4-dimensional space-time, we found that the spectral dimension decrease and become negative as we probe at higher energies.

In this paper, we construct possible modifications to the heat equation given in eqn.(\ref{de}) due to $\kappa$-deformation and study its implication on the scale-dependence of the space-time dimension. For this, we exploit the mapping between the Schr\"odinger equation and heat equation discussed above. Thus, we start from well studied $\kappa$-deformed Klein-Gordon equations written using Beltrami-Laplace operator, in commutative space-time and derive its non-relativistic limit. From thus obtained $\kappa$-deformed Schr\"odinger equation, we construct the deformed heat equation by a Wick's rotation (by implementing the map $t\rightarrow -it$ ). Note that the Wick's rotation is applied to the theory written in the commutative space-time and all the effects of non-commutativity are included through the deformation parameter `$a$' dependent terms\footnote[1]{The Wick's rotation in non-commutative space-time is a non-trivial issue and has been analysed in detail, particularly for the case of moyal space-time in \cite{121, 122, 111}. It was shown in \cite{121} that the naive Wick's rotation will lead to the theory being non-unitary and a consistant way to map non-commutative theory from Euclidean to Minkowski signature was obtained \cite{121, 122, 111, book}.}. Note that the deformation parameter `$a$' is uneffected by the Wick's rotation. This allows us to investigate two related issues, (i) spectral dimension of $\kappa$-deformed space and its scaling with energy (ii) dimensional flow of the full $\kappa$-deformed space-time. The first problem is studied by evaluating the spectral dimension of $\kappa$-space using the $\kappa$-deformed heat equation derived from the $\kappa$-deformed Schr\"odinger equation. Here we use the (d-1) dimensional Laplacian constructed as the space derivative part of the non-relativistic limit of $\kappa$-deformed Klein-Gordon equation, for $\mathcal{L}$ in eqn.(\ref{de}). For investigating the second problem, we take the $\kappa$-deformed heat equation obtained by the Wick's rotation of the $\kappa$-deformed Schr\"odinger equation and replace the Laplacian with the Euclideaniced Beltrami-Laplace operator defined in the d-dimensional $\kappa$-deformed space-time, for $\mathcal{L}$ in eqn.(\ref{de}). We have carried out this study by different choices of $\kappa$-deformed Klein-Gordon equations.

In the first case, thus, we study the spectral dimension of the spatial part of the $\kappa$-deformed space-time. Since the spatial coordinates of the $\kappa$-space-time commute among themselves and it is the time coordinate which do not commute with the space-coordinates, this approach is appropriate to study how the space dimension of the $\kappa$-space-time changes as the probe scale is changed due to the non-commutativity between time and space coordinates. We see that the effect of non-commutativity is to introduce higher spatial derivative terms as well as terms involving both spatial and temporal derivatives in the deformed diffusion equation. The role of these terms on the spectral dimension is brought out here.

In the second case, we take the $\kappa$-deformed heat equation as the starting point of the analysis and replace the Laplacian $\mathcal{L}$ in eqn.(\ref{de}) by the Euclideanised Beltrami-Laplace operator. Thus here, the non-commutativity shows itself in two ways, by introducing the higher derivative terms in the deformed heat equation and also through the additional terms appearing in the deformed Baltrami-Laplace operator. Here also we do the analysis for different choices of Baltrami-Laplace operator.

Organization of this paper is as follows. In the second section, we setup the diffusion equation using the deformed Klein-Gordon equation. We start with the Klein-Gordon equation in $\kappa$-Minkowski space-time written in terms of commuting coordinates and all the effects of non-commutativity are contained in the `$a$'(deformation parameter) dependent terms. By taking the non-relativistic limit of this theory written in terms of the commutative variables and applying Wick's rotation, we derive the diffusion equation in the $\kappa$-deformed Euclidean space, valid upto first non-vanishing terms in the deformation parameter $a$. We then solve this diffusion equation perturbatively and use this solution to calculate the spectral dimension. We have also analyzed the change in the spectral dimension due to extended nature of the probe. In the next subsection, we start with a different choice of generalized $\kappa$-deformed Klein-Gordon equation and arrive at the $\kappa$-deformed diffusion equation. The spectral dimension is calculated using its solution and dimensional flow is analysed. In section 3, we replace the Laplacian in the modified diffusion equation with the two different choices of Beltrami-Laplace operator and use this diffusion equation to calculate the spectral dimension. The analysis of results and summary are presented in the last section.

\section{$\kappa$-deformed diffusion equation and spectral dimension}
In this section, we derive the $\kappa$-deformed diffusion equations starting from two possible choices of $\kappa$-deformed Klein-Gordon equations. The diffusion equation is related to the Schr\"odinger equation under the mapping $t\rightarrow -it $ and we use this map to derive the deformed diffusion equation. By replacing $t$ with $-it$ in the $\kappa$-deformed Schr\"odinger equations, derived by taking the non-relativistic limit of $\kappa$-deformed Klein-Gordon equation, we obtain the $\kappa$-deformed diffusion equations. Using perturbative method, we obtain its solution valid upto second order in the deformation parameter. From this solution, we calculate the return probability which is a measure of finding a particle back at the starting point after a finite time gap. Using this, we calculate the spectral dimension.

\subsection{Diffusion equation from the $\kappa$-deformed Klein-Gordon equation $(D_{\mu}D^{\mu}-m^2)\phi =  0$}
Here we derive the deformed diffusion equation from the non-relativistic limit of $\kappa$-deformed Klein-Gordon equation $(D_{\mu}D^{\mu}-m^2)\phi =  0$. Consider an n-dimensional $\kappa$-deformed Minkowski space with signature (-++...+). The Generalized Klein-Gordon equation on $\kappa$-deformed space-time\cite{kgeqn} is
 \begin{equation}
   \square (1+\frac{a^2}{4} \square )\phi =\frac{m^2 c^2}{\hbar^2}\phi, \label{KGeqn}
 \end{equation}
where
 \begin{equation}
   \square = \nabla_{n-1}^2 \frac{e^{-A}}{\varphi^2}+\partial_{0}^{2}\frac{2(1-coshA)}{A^2}. \label{squr}
 \end{equation}
Here $\nabla_{n-1}^2=\Sigma_{i=1}^{n-1}\frac{\partial^2}{\partial x_i^2}$, $A=-ia\partial_0$, and we choose $\varphi = e^{-\frac{A}{2}}$. We expand eqn.(\ref{KGeqn}) in terms of the deformation parameter and obtain the equation valid upto second-order in $a$ as  
\begin{equation}
   \left(\nabla_{n-1}^2 -\partial_0^2+\frac{a^2}{4}\nabla_{n-1}^4-\frac{a^2}{2}\nabla^2_{n-1}\partial_0^2+\frac{a^2}{3}\partial_0^4\right)\phi = \frac{m^2 c^2}{\hbar^2}\phi. \label{KG}
\end{equation}
We next construct $\kappa$-deformed Schr$\ddot{o}$dinger equation by taking the non-relativistic limit of the $\kappa$-deformed Klein-Gordon equation. Note that the eqn.(\ref{KG}) is written, completly in the commutative space-time. This allows us to use the well known calculation scheme to obtain the non-relativistic limit \cite{ZEE}. Thus we start with the ansatz wave function $\phi$ where one separates out the rest-mass dependence and further we use the fact that in the non-relativistic limit, kinetic energy is very small compared to rest mass energy. So we start with the ansatz
  \begin{equation}
   \phi(x,t)=\varphi(x,t) e^{-i \frac{mc^2}{\hbar}t}\label{nrlimit}
  \end{equation}
in eqn.(\ref{KG}). Here $x$ is a point in the (n-1) space. Effectively, this ansatz allows us to extract a term containing the rest mass $m$. In the non-relativistic limit, the kinetic energy ($KE$) is small compared to rest mass energy, i.e., $KE<<mc^2$ and hence we have
 \begin{equation}
    \mid i\hbar \frac{\partial \varphi}{\partial t} \mid \ll mc^2 \varphi. \label{nr}
 \end{equation}
Substituting eqn.$(\ref{nrlimit})$ in eqn.$(\ref{KG})$ and after using the fact that $KE$ is much smaller than the rest mass energy (stated in eqn.$(\ref{nr})$), we get the $\kappa$-deformed Schr\"odinger equation as
\begin{equation}
\begin{split}
  \nabla_{n-1}^2\varphi+i\frac{2m}{\hbar}\frac{\partial\varphi}{\partial t} +\frac{a^2}{4}\nabla_{n-1}^4\varphi +ia^2\frac{m}{\hbar}\frac{\partial}{\partial t}\nabla^2_{n-1} \varphi + \frac{a^2}{2}\frac{m^2c^2}{\hbar^2}\nabla_{n-1}^2\varphi \\ +ia^2 \frac{4m^3c^2}{3\hbar^3}\frac{\partial\varphi}{\partial t} +a^2 \frac{m^4c^4}{3 \hbar^4} \varphi = 0.
  \end{split}
\end{equation}
By changing $t$ to $-it$ in the above, we get the deformed diffusion equation as
\begin{equation}
\begin{split}
  \nabla_{n-1}^2\varphi-\frac{2m}{\hbar}\frac{\partial\varphi}{\partial t} +\frac{a^2}{4}\nabla_{n-1}^4\varphi -a^2\frac{m}{\hbar}\frac{\partial}{\partial t}\nabla^2_{n-1} \varphi + \frac{a^2}{2}\frac{m^2c^2}{\hbar^2}\nabla_{n-1}^2\varphi  \\ -a^2 \frac{4m^3c^2}{3\hbar^3}\frac{\partial\varphi}{\partial t} +a^2 \frac{m^4c^4}{3 \hbar^4} \varphi = 0. \label{DE}
  \end{split}
\end{equation}
Redefining $kt=\sigma$ with $k=\frac{\hbar}{2m}$ and after some rearrangements, we obtain $\kappa$-deformed diffusion equation as
\begin{equation}
\frac{\partial\varphi}{\partial \sigma}=\nabla_{n-1}^2\varphi+\frac{a^2 c^2}{8 k^2}\nabla_{n-1}^2\varphi+\frac{a^2}{4}\nabla_{n-1}^4\varphi-\frac{a^2}{2}\frac{\partial}{\partial \sigma}\nabla^2_{n-1} \varphi-\frac{a^2 c^2}{6 k^2} \frac{\partial\varphi}{\partial \sigma}+ \frac{a^2 c^4}{48 k^2}  \varphi, \label{diff eqn}
\end{equation}
where $\nabla_{n-1}^2=\Sigma_{i=1}^{n-1}\frac{\partial^2}{\partial x_i^2}$. In the above equation $\varphi$ is a function of $x$ and $\sigma$. It is clear from the eqn.(\ref{diff eqn}) that, in the commutative limit ($a \rightarrow 0 $), we obtain the usual diffusion equation. Note that in deriving $\kappa$-deformed diffusion equation from $\kappa$-deformed Schr\"odinger equation, we only replace t with -it and absorbe $\frac{\hbar}{2m}$ factor into the diffusion scale $\sigma$. The $\kappa$-deformation parameter `$a$' does not get any modification under this mapping. 

Note that the deformed diffusion equation has higher order spatial derivative ($\nabla_{n-1}^4$) and term involving products of temporal and spatial derivatives i.e., $\frac{\partial}{\partial \sigma}\nabla^2_{n-1}$. But there are no higher derivative terms with respect to (scaled) time ($\sigma$). These features would turn out to be significant in the calculation of spectral dimension of $\kappa$-deformed space-time.
\subsection{Spectral dimension}
 To find the heat kernel $\varphi(x,y;\sigma)$ of the $\kappa$-deformed diffusion equation obtained in eqn.(\ref{diff eqn}), we express the solution as a  perturbative series in $a$ as
\begin{equation}
  \varphi=\varphi_0+a\varphi_1+a^2 \varphi_2. \label{pert sol}
\end{equation}
We note that the dimension of the terms satisfy the relations, $[\varphi_1]=\frac{1}{L} [\varphi_0]$ and $[\varphi_2]=\frac{1}{L^2}[\varphi_0]$.
 
 Using eqn.(\ref{pert sol}) in eqn.(\ref{diff eqn}) and equating the terms of same order in $a$, we solve the above equation. The zeroth order terms in $a$ leads to
 \begin{equation}
   \frac{\partial}{\partial \sigma} \varphi_0(x,y;\sigma) = \nabla^2_{n-1} \varphi_0(x,y;\sigma).
 \end{equation}
The laplacian $\nabla^2_{n-1}$ is with respect to $x$ coordinates and will act on the x dependence of the heat kernel. The solution to this equation is given by 
  \begin{equation}
   \varphi_0(x,y;\sigma)=\frac{1}{(4\pi \sigma)^\frac{n-1}{2}} e^{-\frac{\Sigma_{i=1}^{n-1}(x_i-y_i)^2 }{4\sigma}}. \label{psi0}
 \end{equation}
Next, equating the first order terms in $a$ gives us the equation
  \begin{equation}
   \frac{\partial}{\partial \sigma} \varphi_1(x,y;\sigma) = \nabla^2_{n-1} \varphi_1(x,y;\sigma). \label{eqn2}
 \end{equation} 
Note that here too, $\nabla^2_{n-1}$ is the Laplacian with respect to $x$ coordinates (and this notation is used in the remaining part of this paper) and this will act on the first argument of $\varphi_1$, namely $x$.
 
 The solution $\varphi_1(x,y;\sigma)$  satisfying the above equation also have the same form as $\varphi_0(x,y;\sigma)$ since both satisfy the same heat equation. Thus we get
  \begin{equation}
   \varphi_1(x,y;\sigma)=\frac{\alpha}{(4\pi \sigma)^\frac{n-1}{2}} e^{-\frac{\Sigma_{i=1}^{n-1}(x_i-y_i)^2}{4\sigma}}. \label{psi1}
 \end{equation}
where the constant $\alpha$ has dimension of $L^{-1}$. Now by equating the second order terms in $a$ in eqn.(\ref{diff eqn}), we find
\begin{equation}
  \frac{\partial\varphi_2}{\partial \sigma}=\nabla_{n-1}^2\varphi_2+\frac{c^2}{8 k^2}\nabla_{n-1}^2\varphi_0+\frac{1}{4}\nabla_{n-1}^4\varphi_0-\frac{1}{2}\frac{\partial}{\partial \sigma}\nabla^2_{n-1} \varphi_0-\frac{c^2}{6 k^2} \frac{\partial\varphi_0}{\partial \sigma}+ \frac{c^4}{48 k^2}  \varphi_0.
\end{equation}
Substituting the solution for $\varphi_0$ from eqn.(\ref{psi0}) in the above equation and after straight forward manipulations, we get 
\begin{equation}
\begin{split}
\frac{\partial\varphi_2}{\partial\sigma} = \nabla_{n-1}^2\varphi_2+ [\frac{c^4}{48 k^4}+\frac{c^2}{48 k^2}\frac{(n-1)}{\sigma}-\frac{(n^2-1)}{16 \sigma^2}-\frac{c^2}{96 k^2}\frac{\Sigma_{i=1}^{n-1}(x_i-y_i)^2}{\sigma^2}\\+\frac{(n+1)}{16 \sigma^3}\Sigma_{i=1}^{n-1}(x_i-y_i)^2-\frac{1}{64 \sigma^4}(\Sigma_{i=1}^{n-1}(x_i-y_i)^2)^2 ]  \frac{1}{(4\pi \sigma)^\frac{n-1}{2}} e^{-\frac{\Sigma_{i=1}^{n-1}(x_i-y_i)^2 }{4\sigma}}. 
\end{split}
\end{equation}
The above equation is of the generic form 
\begin{equation}
  \frac{\partial}{\partial \sigma}\varphi_2(X,\sigma)=\nabla^2_{n-1}\varphi_2(X,\sigma)+f(X,\sigma).
\end{equation}
For a given initial condition, $ \varphi_2(X,0) = g(X)$, the solution to above equation
can be written as\cite{duhamel}
\begin{equation}
  \varphi_2(X,\sigma)=\int_{R^{n-1}} \Phi (X-X',\sigma)g(X')dX' + \int_{0}^{\sigma}\int_{R^{n-1}} \Phi (X-X',\sigma-s) f(X',s) dX' ds,\label{phi2}
\end{equation}
where 
\begin{equation}
  \Phi(X,\sigma)=\frac{1}{(4\pi \sigma)^\frac{n-1}{2}} e^{-\frac{\left(X_1^2+X_2^2+...+X_{n-1}^2\right)}{4\sigma}}.
\end{equation}
Using the initial condition $\varphi_2(X,0)=\delta^{n-1}(X)$, we obtain the first term $\varphi_{21}$ in the eqn.(\ref{phi2}) as
\begin{equation}
  \varphi_{21}(x,y;\sigma) = \frac{\beta}{(4\pi \sigma)^\frac{n-1}{2}} e^{-\frac{\Sigma_{i=1}^{n-1}(x_i-y_i)^2}{4\sigma}}\label{psi21}
\end{equation}
where $\beta$ has dimension $L^{-2}$. The second term on RHS of eqn.$(\ref{phi2}$), $\varphi_{22}$ is calculated as

\begin{eqnarray}
  \varphi_{22}(x,y;\sigma) &=&  \frac{1}{(4\pi \sigma)^\frac{n-1}{2}} e^{-\frac{\Sigma_{i=1}^{n-1}(x_i-y_i)^2 }{4\sigma}} [\left(\frac{c^4}{48k^4} - \frac{n^2-1}{16 \sigma^2} + \frac{c^2}{48 k^2} \frac{(n-1)}{\sigma}\right)(\sigma-\epsilon) \nonumber \\ & - & \left(\frac{c^2}{96k^2\sigma^2} + \frac{1}{64\sigma^4}\Sigma_{i=1}^{n-1}(x_i-y_i)^2 -\frac{(n+1)}{16 \sigma^3}\right)\Sigma_{i=1}^{n-1}(x_i-y_i)^2(\sigma-\epsilon) \nonumber \\ & - & \left(\frac{c^2}{24k^2}\frac{1}{\sigma\sqrt{\sigma\pi}}+\frac{1}{8\sigma^3\sqrt{\sigma\pi}}\Sigma_{i=1}^{n-1}(x_i-y_i)^2\right)\Sigma_{i=1}^{n-1}(x_i-y_i) (\sigma \tan^{-1}q-\epsilon q) \nonumber \\ &-& \frac{1}{2\sigma^3\pi}\left[(x_1-y_1)\Sigma_{i=2}^{n-1}(x_i-y_i)+(x_2-y_2)\Sigma_{i=3}^{n-1}(x_i-y_i)+...+(x_{n-2}-y_{n-2})(x_{n-1}-y_{n-1})\right] A \nonumber \\ & + & \frac{1}{4\sigma^2\sqrt{\sigma\pi}}\Sigma_{i=1}^{n-1}(x_i-y_i)\left((5n+2)\sigma \tan^{-1}q-(4n+2)\sigma q -nq\epsilon\right)] \label{psi22}
\end{eqnarray}
Where $q = \sqrt{\frac{\sigma}{\epsilon}-1}$ and $ A = \sigma \ln(\sigma/\epsilon) - \sigma + \epsilon) $. 
Using  eqns.$(\ref{psi0}),(\ref{psi1}),(\ref{psi21})$ and $(\ref{psi22})$ in eqn.$(\ref{pert sol})$, we find the heat kernel valid upto second order in $a$. Using the definition (eqn.(\ref{p_g})) of return probability we obtain $P(\sigma)$ (in the limit $\epsilon\rightarrow 0$) as
\begin{equation}
 P(\sigma) = \frac{1}{(4\pi\sigma)^{\frac{n-1}{2}}}\left[1+a\alpha+a^2\beta+a^2\frac{c^4}{48k^4}\sigma - a^2\frac{(n^2-1)}{16\sigma}+a^2\frac{c^2}{48k^2}(n-1)\right].
\end{equation}
The spectral dimension is found by taking the logarithmic derivative of the above return probability. Thus we get
\begin{equation}
  D_s = (n-1) - \frac{a^2}{8}\frac{(n^2-1)}{\sigma}-\frac{a^2 c^4}{24k^4}\sigma.\label{spedim}
\end{equation}
From the above expression, we see that apart from the usual (n-1) term, we have two additional terms and both of them are of second order in $a$. They arise due to the non-commutative nature of $\kappa$-space-time. One term is dependent on the topological dimension n we started with and the other term is independent of the initial dimension. Note that the diffusion scale $\sigma$ appear in the n dependent correction as inverse where as in the second correction term, it appears linearly. In the commutative limit, we see that the spectral dimension is same as the topological dimension i.e $D_s = n-1$.

\begin{figure}
\caption{ spectral dimension as a function of $\sigma$ for n=4, $a$=1, c=k=1.}\label{fig1}
\includegraphics[width=3in]{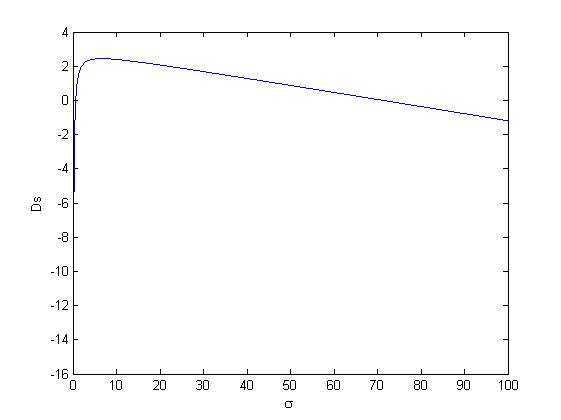}
\end{figure}

For n=4 with a=c=k=1, we obtain $D_s = 3-\frac{\sigma}{24}-\frac{15}{8 \sigma}$. From this, we see that in the limit $\sigma \rightarrow $ 0, the spectral dimension $D_s \rightarrow -\infty$. As $\sigma$ increases the spectral dimension also increases and reaches a maximum value $D_s \sim 2.44$ for $\sigma \sim 6.7$. As we go further, the spectral dimension start decreasing [see fig.(\ref{fig1})].

In general, for n=4 and with c=k=1, we get an inequality for $\sigma$,  $\frac{36}{a^2}-\sqrt{\frac{1296}{a^4}-45} < \sigma <\frac{36}{a^2}+\sqrt{\frac{1296}{a^4}-45}$ where the spectral dimension become positive and it takes negative value outside this range. The condition on the deformation parameter, $a^2 < \frac{72 \sigma}{45+\sigma^2}$, implies that the spectral dimension is positive.

We also investigate the effect of the extended nature of the probe on the spectral dimension. For this purpose, we consider Gaussian distribution as our initial condition in solving eqn.(\ref{diff eqn}), i.e., we take
\begin{equation}
\varphi(x,y;0)=\frac{1}{(4 \pi a^2)^{(\frac{n-1}{2})}} e^{-\frac{\Sigma_{i=1}^{n-1}(x_i-y_i)^2}{4 a^2}},
\end{equation}
instead of the delta function condition used to obtain $\varphi_0, \varphi_1$ and $\varphi_2 $. Using this, we solve eqn.(\ref{diff eqn}) and obtain the zeroth order solution as 
\begin{equation}
\varphi_0(x,y;\sigma)= \frac{1}{(4\pi(\sigma+a^2))^{(\frac{n-1}{2})}}e^{- \frac{\Sigma_{i=1}^{n-1}(x_i-y_i)^2}{4(\sigma+a^2)}}.
\end{equation}
Keeping terms upto second order in $a$, we find
\begin{equation}
\varphi_0(x,y;\sigma)= \frac{1}{(4\pi\sigma)^{(\frac{n-1}{2})}}e^{- \frac{\Sigma_{i=1}^{n-1}(x_i-y_i)^2}{4\sigma}}\left(1+\frac{a^2}{4\sigma^2}\Sigma_{i=1}^{n-1}(x_i-y_i)^2-(n-1)\frac{a^2}{2\sigma}\right).
\end{equation}
Similarly we obtain $\varphi_1$ from eqn.(\ref{eqn2}), valid up to first order in $a$ as
\begin{equation}
\varphi_1(x,y;\sigma)= \frac{\alpha}{(4\pi\sigma)^{(\frac{n-1}{2})}}e^{- \frac{\Sigma_{i=1}^{n-1}(x_i-y_i)^2}{4\sigma}}.
\end{equation}
For $\varphi_2$ we need to consider only the zeroth order terms in $a$, since expression for $\varphi$ contains $a^2\varphi_2$, and thus the solution for $\varphi_2$ will be the same as we obtained in eqns.(\ref{psi21}) and (\ref{psi22}).

Using this we calculate the spectral dimension as
\begin{equation}
D_s = (n-1)- \frac{a^2 c^4}{24k^4}\sigma - \frac{a^2}{8}\frac{(n^2-1)}{\sigma}-\frac{a^2}{\sigma} (n-1).
\end{equation}
By comparing with the eqn.(\ref{spedim}), we note that we have an extra term $-\frac{a^2}{\sigma} (n-1)$, which is due to the extended nature of the probe. Further, we note that the dimensional flow has the same general behaviour as the one obtained with point particle probe in eqn.(\ref{spedim}). Here again, we note that there are terms with $\sigma^{-1}$ dependence and one term with linear dependence on $\sigma$, the diffusion scale. The finite size effect of the test particle introduces a correction which is proportional to the inverse power of $\sigma$.

\subsection{Diffusion equation for $(\square-m^2)\phi=0$ and spectral dimension}
Eqn.(\ref{KGeqn}) and eqn.(\ref{squr}) show that both $D_{\mu} D^{\mu}$ and $\square$ operator have the same commutative limit. Thus, the requirement of correct commutative limit allow
\begin{equation}
   \square \phi =\frac{m^2 c^2}{\hbar^2}\phi. \label{box}
\end{equation}
as a possible $\kappa$-deformed Klein-Gordon equation. Expanding this equation up to first non-vanishing terms in $a$, we find 
\begin{equation}
 \left( \nabla_{n-1}^2-\partial_0^2+\frac{a^2}{12}\partial_0^4\right)\phi = \frac{m^2 c^2}{\hbar^2}\phi. \label{expbox}
\end{equation}
 Using eqn.(\ref{nrlimit}) and eqn.(\ref{nr}) in the above, we obtain the non-relativistic limit of eqn.(\ref{expbox}) as
\begin{equation}
  \nabla_{n-1}^2 \varphi +i\frac{2m}{\hbar}\frac{\partial \varphi}{\partial t}+i\frac{a^2}{3}\frac{m^3c^2}{\hbar^3}\frac{\partial \varphi}{\partial t}+\frac{a^2}{12}\frac{m^4c^4}{\hbar^4}\varphi=0.
\end{equation}
After mapping $t$ to $-it$ and redefining $kt=\sigma$ (where $k=\frac{\hbar}{2m})$ we re-express the above equation as 
\begin{equation}
  \frac{\partial\varphi}{\partial\sigma}=\nabla_{n-1}^2 \varphi -\frac{a^2 c^2}{24k^2}\frac{\partial\varphi}{\partial\sigma}+a^2\frac{c^4}{192k^4}\varphi. \label{deqnsq}
\end{equation}
Unlike eqn.(\ref{diff eqn}), here we do not have higher derivatives terms. We perturbatively solve this deformed diffusion equation using the series expansion of $\varphi$ given in eqn.(\ref{pert sol}). The zeroth order terms gives
\begin{equation}
   \frac{\partial}{\partial \sigma} \varphi_0(x,y;\sigma)=\nabla^2_{n-1} \varphi_0(x,y;\sigma),
 \end{equation}
whose solution is  
 \begin{equation}
   \varphi_0(x,y;\sigma)=\frac{1}{(4\pi \sigma)^\frac{n-1}{2}} e^{-\frac{\Sigma_{i=1}^{n-1}(x_i-y_i)^2 }{4\sigma}}. \label{apsi0}
 \end{equation}
Equating the first order terms in $a$ on both sides of eqn.(\ref{deqnsq}) gives
  \begin{equation}
   \frac{\partial}{\partial \sigma} \varphi_1(x,y;\sigma)=\nabla^2_{n-1} \varphi_1(x,y;\sigma). \label{eqnpsi1}
 \end{equation} 
 The solution to this equation is   
 \begin{equation}
   \varphi_1(x,y;\sigma)=\frac{\alpha}{(4\pi \sigma)^\frac{n-1}{2}} e^{-\frac{\Sigma_{i=1}^{n-1}(x_i-y_i)^2}{4\sigma}}.\label{apsi1}
 \end{equation} 
Note $\alpha$ has the dimension of inverse length. Next we collect terms of having $a^2$ from both 
sides of eqn.(\ref{deqnsq}) to get
\begin{equation}
  \frac{\partial\varphi_2}{\partial\sigma}=\nabla_{n-1}^2 \varphi_2 -\frac{c^2}{24k^2}\frac{\partial\varphi_0}{\partial\sigma}+\frac{c^4}{192k^4}\varphi_0. \label{aphi2}
\end{equation}
Substituting for $\varphi_0$ from eqn.(\ref{apsi0}) in the above, reduces eqn.(\ref{aphi2}) to 
\begin{equation}
  \frac{\partial\varphi_2}{\partial\sigma}=\nabla_{n-1}^2 \varphi_2+\left[\frac{c^2}{48k^2}\frac{(n-1)}{\sigma}-\frac{c^2}{96k^2}\frac{1}{\sigma^2}\Sigma_{i=1}^{n-1}(x_i-y_i)^2+\frac{c^4}{192k^4}\right] \frac{1}{(4\pi \sigma)^\frac{n-1}{2}} e^{-\frac{\Sigma_{i=1}^{n-1}(x_i-y_i)^2}{4\sigma}}.\label{aaphi2}
\end{equation}
Using eqn.(\ref{phi2}), we solve this differential equation. Then the first term of eqn.(\ref{phi2}) will give $\varphi_{21}$ as
\begin{equation}
   \varphi_{21}(x,y;\sigma) = \frac{\beta}{(4\pi \sigma)^\frac{n-1}{2}} e^{-\frac{\Sigma_{i=1}^{n-1}(x_i-y_i)^2 }{4\sigma}}\label{apsi21}
\end{equation}
where $\beta$ has dimension $L^{-2}$. The second term on RHS of eqn.(\ref{phi2}), $\varphi_{22}$ is evaluated as
\begin{equation}
\begin{split}
  \varphi_{22}(x,y;\sigma) = \frac{1}{(4\pi \sigma)^\frac{n-1}{2}} e^{-\frac{\Sigma_{i=1}^{n-1}(x_i-y_i)^2}{4\sigma}}[\left(\frac{c^4}{192k^4}-\frac{c^2}{96k^2\sigma^2}\Sigma_{i=1}^{n-1}(x_i-y_i)^2+\frac{c^2}{48k^2\sigma}(n-1)\right)(\sigma-\epsilon) \\-\frac{c^2}{24k^2}\frac{1}{\sigma \sqrt{\sigma \pi}}\Sigma_{i=1}^{n-1}(x_i-y_i)^2 \left(\sigma \tan^{-1}\sqrt{\frac{\sigma}{\epsilon}-1}-\epsilon\sqrt{\frac{\sigma}{\epsilon}-1}\right)]. \label{apsi22}
\end{split}
\end{equation}
Using  eqns.(\ref{apsi0}),(\ref{apsi1}),(\ref{apsi21}) and eqn.(\ref{apsi22}) in eqn.(\ref{pert sol}) we find the heat kernel valid upto second order in $a$. From this we calculate the return probability  (in the limit $\epsilon\rightarrow$ 0) as
\begin{equation}
 P(\sigma) = \frac{1}{(4\pi\sigma)^{\frac{n-1}{2}}}\left[1+a\alpha+a^2\beta+a^2\frac{c^4}{192k^4}\sigma+a^2\frac{c^2}{48k^2}(n-1)\right].
\end{equation}
Using this we find the Spectral dimension to be
\begin{equation}
  D_s = (n-1)- \frac{a^2 c^4}{96k^4}\sigma. \label{sdsq}
\end{equation}

\begin{figure}
\caption{ spectral dimension as a function of $\sigma$ for n=4, $a$=1, c=k=1.}\label{fig2}
 \includegraphics[width=3in]{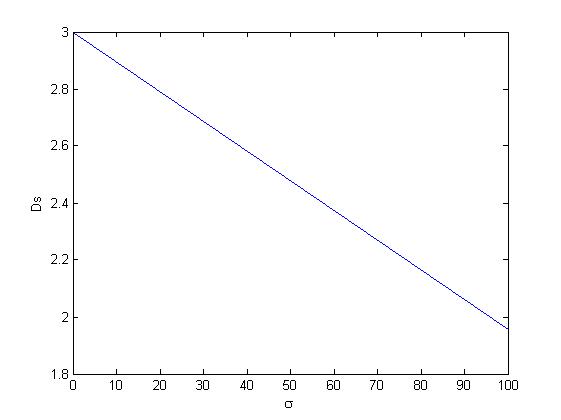}
\end{figure}

The correction of the spectral dimension is of second order in $a$ and it is independent of the initial dimension. Thus we see that the change in spectral dimension is same for space-times of all dimensions. Here we see that the $a$ dependent correction to the spectral dimension is linear in the diffusion scale $\sigma$. Unlike the spectral dimension obtained in eqn.(\ref{spedim}), there is no term involving $\sigma^{-1}$ in eqn.(\ref{sdsq}).

In the commutative limit we have $D_s=n-1$, same as the topological dimension. For n=4, k=c=1, it is easy to see from the Fig.\ref{fig2} that spectral dimension $D_s=3$ exactly at $\sigma = 0$, and it start decreasing as $\sigma$ increases. For $\sigma=\frac{288}{a^2}$ the spectral dimension vanishes and it is negative for higher values of $\sigma$.

Now we want to see the change in spectral dimension due to the extended nature of probe. We use Gaussian function as initial condition and solve for the heat kernel. The modified initial condition will be
\begin{equation}
\varphi(x,y;0)=\frac{1}{(4 \pi a^2)^{(\frac{n-1}{2})}} e^{- \frac{\Sigma_{i=1}^{n-1}(x_i-y_i)^2}{4 a^2}}.
\end{equation}
Using this initial condition, we solve eqn.(\ref{deqnsq}) and obtain the zeroth order term as
\begin{equation}
\varphi_0(x,y;\sigma)= \frac{1}{\left[4\pi(\sigma+a^2)\right]^{(\frac{n-1}{2})}}e^{- \frac{\Sigma_{i=1}^{n-1}(x_i-y_i)^2}{4(\sigma+a^2)}}.
\end{equation}
Since we are interested only upto second order terms in $a$, we expand this as 
\begin{equation}
\varphi_0(x,y;\sigma)= \frac{1}{(4\pi\sigma)^{(\frac{n-1}{2})}}e^{- \frac{\Sigma_{i=1}^{n-1}(x_i-y_i)^2}{4\sigma}}\left(1+\frac{a^2}{4\sigma^2}\Sigma_{i=1}^{n-1}(x_i-y_i)^2-(n-1)\frac{a^2}{2\sigma}\right).
\end{equation}
Similarly we obtain $\varphi_1$ from eqn.(\ref{eqnpsi1}), valid upto first order in $a$ as
\begin{equation}
\varphi_1(x,y;\sigma)= \frac{\alpha}{(4\pi\sigma)^{(\frac{n-1}{2})}}e^{- \frac{\Sigma_{i=1}^{n-1}(x_i-y_i)^2}{4\sigma}}.
\end{equation}
The equation for $\varphi_2$ will be same as eqn.(\ref{aaphi2}), since we are interested only in terms of the order $a^2$. The resultant solution will be same as eqns.(\ref{apsi21}) and (\ref{apsi22}). Using this we obtain the spectral dimension as
\begin{equation}
 D_s = (n-1)- \frac{a^2 c^4}{96k^4}\sigma - (n-1)\frac{a^2}{\sigma}.
\end{equation}
By comparing with the eqn.(\ref{sdsq}), here we have an extra term $-\frac{a^2}{\sigma} (n-1)$ due to the extended nature of the probe. Thus we see that the extended nature of the probe introduce a correction to the spectral dimension which depends on the inverse power of the diffusion scale $\sigma$.

\section{Modified $\kappa$-diffusion equation and spectral dimension}
In this section, we study alternative diffusion equations than the ones analysed in the previous section. Here, we generalise the approach where one starts from the diffusion equation and replace the Laplacian ($\mathcal{L}$ in eqn.(\ref{de})) with the Beltrami-Laplace operator. Thus, we start with the $\kappa$-deformed diffusion equation derived in eqn.(\ref{diff eqn}), but use the $\kappa$-deformed Beltrami-Laplace operator in place of the Laplacian $\nabla^2_{n-1}$, keeping all other terms of eqn.(\ref{diff eqn}) unchanged. Thus in this approach, we include the possible modification of the diffusion equation in $\kappa$-space-time coming from two sources. First due to the additional terms in the diffusion equation involving the derivative with respect to the diffusion time $\sigma$, and second due to the non-local and higher derivatives terms appearing through the deformed Beltrami-Laplace operator. As earlier, here too we analyse the spectral dimension using two different choices of $\kappa$-deformed Beltrami-Laplace operator.
\subsection{Diffusion equation with Beltrami-Laplace operator and corresponding spectral dimension}
In this subsection, we rewrite the diffusion equation eqn.(\ref{diff eqn}) using the Casimir (general form of Laplacian) of the kappa-Euclidean space. The Casimir of d-dimensional $\kappa$-deformed Euclidean space is given by\cite{sm,17a,17b} 
\begin{eqnarray}
 &D_{\mu}D_{\mu}=\square (1-\frac{a^2}{4} \square ) \label{dmu} \\&
  \square = \nabla^2_{d-1} \frac{e^{-A}}{\varphi^2}-\partial_{d}^{2}\frac{2(1-coshA)}{A^2}\label{square}
\end{eqnarray}
where $\nabla^2_{d-1}=\Sigma_{i=1}^{d-1}\frac{\partial^2}{\partial x_i^2}$ and $\partial_{d}^{2} = \frac{\partial^2}{\partial x_d^2}$. Here $x_d$ is the Euclidean time coordinate and $x_i$, $i=1,2,...,d-1$ are the space coordinates.

Eqn.(\ref{diff eqn}), for a generic n-dimensional Euclidean space reads as 
\begin{equation}
\frac{\partial\varphi}{\partial \sigma}=\nabla_n^2\varphi+\frac{a^2 c^2}{8 k^2}\nabla_n^2\varphi+\frac{a^2}{4}\nabla_n^4\varphi-\frac{a^2}{2}\frac{\partial}{\partial \sigma}\nabla^2_n \varphi-\frac{a^2 c^2}{6 k^2} \frac{\partial\varphi}{\partial \sigma}+ \frac{a^2 c^4}{48 k^2}  \varphi. \label{d diff eqn} 
\end{equation}
Since the above equation is valid for any dimensions, we use $D_{\mu}D_{\mu}$ for $\nabla_n^2$, which is the general form of the Beltrami-Laplace operator in the $\kappa$-deformed Euclidean space. 

We expand eqn.(\ref{dmu}) upto first non-vanishing terms in $a$,
\begin{equation}
 D_{\mu}D_{\mu} = \nabla_{d-1}^2 + \partial_d^2-\frac{a^2}{3} \partial_d ^4  -\frac{a^2}{2}\nabla_{d-1}^2 \partial_d ^2 - \frac{a^2}{4} \nabla_{d-1}^4, \label{dmuex}
\end{equation}
and use this in eqn.(\ref{d diff eqn}) and keep terms upto second order in $a$
\begin{equation}
\begin{split} 
\frac{\partial\varphi}{\partial \sigma}=\nabla_{n-1} ^2 \varphi + \partial_n^2 \varphi-\frac{a^2}{3} \partial_n ^4\varphi  -\frac{a^2}{2}\nabla_{n-1}^2 \partial_n ^2\varphi - \frac{a^2}{4} \nabla_{n-1}^4\varphi+\frac{a^2 c^2}{8 k^2}\left[\nabla_{n-1}^2\varphi+\partial_n^2 \varphi \right]\\ +\frac{a^2}{4}\left[\nabla_{n-1}^4\varphi+\partial_n^4\varphi+2\nabla_{n-1}^2 \partial_n^2 \varphi\right]-\frac{a^2}{2}\frac{\partial}{\partial \sigma}\left[\nabla^2_{n-1} \varphi+\partial_n^2 \varphi\right]-\frac{a^2 c^2}{6 k^2} \frac{\partial\varphi}{\partial \sigma}+ \frac{a^2 c^4}{48 k^2}\varphi. \label{eqndiff}
\end{split}
\end{equation}
By comparing with eqn.(\ref{d diff eqn}), we see that there are three extra terms in the above equation and they modify the spectral dimension (obtained in section 2.2). Note that the extra terms  are of higher derivatives in space and Euclidean time coordinates. We have a term which is quartic derivative in Euclidean time, terms involving product of derivatives in space and Euclidean time and a term having quartic derivatives in space coordinate. We solve the above diffusion equation perturbatively using eqn.(\ref{pert sol}) for $\varphi$. By equating the zeroth order terms in $a$ we obtain
\begin{equation}
   \frac{\partial}{\partial \sigma} \varphi_0(x,y;\sigma)=\nabla^2_{n-1} \varphi_0(x,y;\sigma)+ \partial_n^2\varphi_0(x,y;\sigma).
 \end{equation}
This is the usual diffusion equation in n-dimension whose solution is  
 \begin{equation}
   \varphi_0(x,y;\sigma)=\frac{1}{(4\pi \sigma)^\frac{n}{2}} e^{-\frac{\mid x-y \mid^2}{4\sigma}}. 
 \end{equation}
The first order term in $a$ will give 
 \begin{equation}
  \frac{\partial}{\partial \sigma} \varphi_1(x,y;\sigma)=\nabla^2_{n-1} \varphi_1(x,y;\sigma)+ \partial_n^2\varphi_1(x,y;\sigma).
 \end{equation} 
It is clear that $\varphi_1(x,y;\sigma)$ also satisfy the usual heat equation and thus  
 \begin{equation}
   \varphi_1(x,y;\sigma)=\frac{\alpha}{(4\pi \sigma)^\frac{n}{2}} e^{-\frac{\mid x-y \mid^2}{4\sigma}}.
 \end{equation} 
Now equate the second order terms in $a$ in eqn.(\ref{eqndiff}) to get
\begin{equation}
\begin{split} 
\frac{\partial\varphi_2}{\partial \sigma}=\nabla_{n-1} ^2 \varphi_2 + \partial_n^2 \varphi_2-\frac{1}{3} \partial_n ^4\varphi_0  -\frac{1}{2}\nabla_{n-1}^2 \partial_n ^2\varphi_0 - \frac{1}{4} \nabla_{n-1}^4\varphi_0+\frac{c^2}{8 k^2}\left(\nabla_{n-1}^2\varphi_0+\partial_n^2 \varphi_0\right)\\ +\frac{1}{4}\left(\nabla_{n-1}^4\varphi_0+\partial_n^4\varphi_0+2\nabla_{n-1}^2 \partial_n^2 \varphi_0\right)-\frac{1}{2}\frac{\partial}{\partial \sigma}\left(\nabla^2_{n-1} \varphi_0+\partial_n^2 \varphi_0\right)-\frac{c^2}{6 k^2} \frac{\partial\varphi_0}{\partial \sigma}+ \frac{c^4}{48 k^2}\varphi_0.
\end{split}
\end{equation}
Substitute for $\varphi_0$ and using eqn.(\ref{phi2}) we calculate $\varphi_2$. Using this heat kernel, we obtain the return probability as
\begin{equation}
 P(\sigma) = \frac{1}{(4\pi\sigma)^{\frac{n}{2}}}\left[1+a\alpha+a^2\beta-(1+n)^2\frac{a^2}{16\sigma}+a^2\frac{c^4}{48k^4}\sigma-a^2\frac{n(n+2)}{16\sigma}+\frac{a^2c^2}{48k^2}n\right].
\end{equation}
The logarithmic derivative of the above expression give the spectral dimension as
\begin{equation}
 D_s = n-\frac{a^2}{8\sigma}(1+4n+2n^2)-\frac{a^2 c^4}{24 k^4}\sigma. \label{sd3}
\end{equation}

\begin{figure}
\caption{ spectral dimension as a function of $\sigma$ for n=4, $a$=1, c=k=1.}\label{fig3}
 \includegraphics[width=3in]{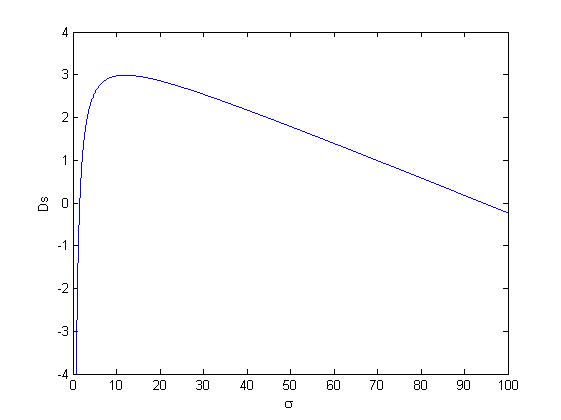}
\end{figure}
In the commutative limit we find $D_s = n$. Spectral dimension as a function of $\sigma$ with n=4 and a=k=c=1 is shown in the Fig.\ref{fig3}. We see that in the limit $\sigma \rightarrow $ 0, the spectral dimension $D_s \rightarrow -\infty$. As $\sigma$ increases, $D_s$ reaches a value close to 3 and thereafter decreases with increase in $\sigma$. We note that one of the correction depend on the diffusion scale linearly while the other changes as the inverse of $\sigma$. This feature is same as the spectral dimension obtained in eqn.(\ref{spedim}). The requirement of the positivity of spectral dimension gives a bound on the deformation parameter as $a^2 < \frac{96 \sigma}{147+ \sigma^2}$.

The use of an extended probe would result the spectral dimension 
\begin{equation}
 D_s = n-\frac{a^2}{8\sigma}(1+4n+2n^2)-\frac{a^2 c^4}{24 k^4}\sigma-\frac{a^2 n}{\sigma}.
\end{equation}
By comparing with eqn.(\ref{sd3}) we find an additional term $-\frac{a^2 n}{\sigma}$ due to the finite width of the probe. This new term is proportional to the initial dimension we start with and inversely proportional to $\sigma$. Note that the extended probe does not change the generic behaviour of the dimensional flow.
\subsection{Spectral dimension with $\square$ as the Beltrami-Laplace operator}
It is easy to see from eqn.(\ref{dmu}) and eqn.(\ref{square}) that the $\square$ operator has the same commutative limit as $D_{\mu} D_{\mu}$. The eqn.(\ref{deqnsq}) in generic n-dimension space-time is of the form  
\begin{equation}
  \frac{\partial\varphi}{\partial\sigma}=\nabla_n^2 \varphi -\frac{a^2 c^2}{24k^2}\frac{\partial\varphi}{\partial\sigma}+a^2\frac{c^4}{192k^4}\varphi. \label{ddeqnsq}
\end{equation}
Now we use $\square$ as the general form of Beltrami-Laplace operator in the above equation, in place of $\nabla_n^2$. We expand the $\square$ operator and keep terms upto first non-vanishing terms in $a$, 
\begin{equation}
\square = \nabla_{d-1}^2+\partial_d^2-\frac{a^2}{12}\partial_d^4.\label{square1}
\end{equation}
Now substituting eqn.(\ref{square1}) in eqn.(\ref{ddeqnsq}) and keeping terms upto second order in $a$, we get
\begin{equation}
  \frac{\partial\varphi}{\partial\sigma}=\nabla_{n-1}^2 \varphi +\partial_n^2 \varphi-\frac{a^2}{12}\partial_n^4\varphi -\frac{a^2 c^2}{24k^2}\frac{\partial\varphi}{\partial\sigma}+a^2\frac{c^4}{192k^4}\varphi.\label{boxdiff}
\end{equation}
We note that eqn.(\ref{boxdiff}) has one extra term compared to eqn.(\ref{ddeqnsq}) which is quartic derivative in the Euclidean time.
We solve the above diffusion equation perturbatively using eqn.(\ref{pert sol}) for $\varphi$, as earlier. By equating the zeroth order terms in $a$ we obtain
\begin{equation}
   \frac{\partial}{\partial \sigma} \varphi_0(x,y;\sigma)=\nabla^2_{n-1} \varphi_0(x,y;\sigma)+ \partial_n^2 \varphi_0(x,y;\sigma) \label{0}
 \end{equation}
and corresponding solution is  
 \begin{equation}
   \varphi_0(x,y;\sigma)=\frac{1}{(4\pi \sigma)^\frac{n}{2}} e^{-\frac{\mid x-y \mid^2}{4\sigma}}. 
 \end{equation}
The first order terms in $a$ give
 \begin{equation}
  \frac{\partial}{\partial \sigma} \varphi_1(x,y;\sigma)=\nabla^2_{n-1} \varphi_1(x,y;\sigma)+\partial_n^2 \varphi_1(x,y;\sigma)\label{1}
 \end{equation} 
 whose solution is given by   
 \begin{equation}
   \varphi_1(x,y;\sigma)=\frac{\alpha}{(4\pi \sigma)^\frac{n}{2}} e^{-\frac{\mid x-y \mid^2}{4\sigma}}.
 \end{equation} 
Second order terms in $a$ will result in
\begin{equation}
  \frac{\partial\varphi_2}{\partial\sigma}=\nabla_{n-1}^2 \varphi_2+ \partial_n^2\varphi_2-\frac{1}{12}\partial_n^4\varphi_0 -\frac{c^2}{24k^2}\frac{\partial\varphi_0}{\partial\sigma}+\frac{c^4}{192k^4}\varphi_0.\label{2}
\end{equation}
We solve this equation by substituting for $\varphi_0$ and using eqn.(\ref{phi2}). The solutions of eqn.(\ref{0}, \ref{1}) and eqn.(\ref{2}) are used to obtain the return probability,
\begin{equation}
P(\sigma) = \frac{1}{(4\pi \sigma)^\frac{n}{2}}\left[1+a\alpha+a^2\beta -\frac{a^2}{16 \sigma}+\frac{a^2 c^4}{192 k^4}\sigma + \frac{a^2 c^2}{48 k^2}n \right],
\end{equation}
and using eqn.(\ref{d_s}), we find the spectral dimension as
\begin{equation}
 D_s = n-\frac{a^2}{8\sigma}-\frac{a^2 c^4}{96 k^4}\sigma. \label{sd4}
\end{equation}
\begin{figure}
\caption{ spectral dimension as a function of $\sigma$ for n=4, $a$=1, c=k=1.}\label{fig4}
 \includegraphics[width=3in]{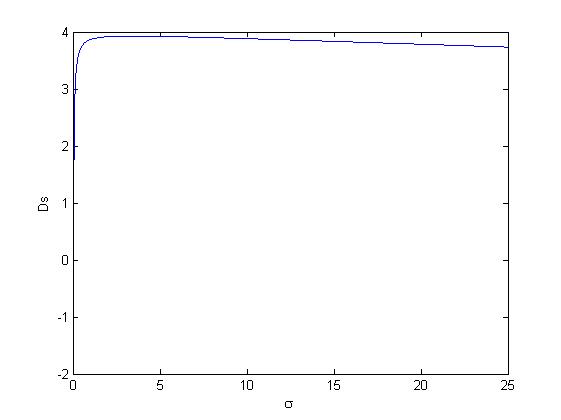}
\end{figure}
Note that the spectral dimension has one term which is linear in $\sigma$ and another which is proportional to $\sigma^{-1}$. For n=4 and k=c=1, it is easy to see from the Fig.\ref{fig4} that, the spectral dimension increases with $\sigma$ initially and then decreases as $\sigma$ increases. It is clear that the spectral dimension is positive for $\frac{192}{a^2}-\sqrt{\frac{36864}{a^4}-12} < \sigma < \frac{192}{a^2}+\sqrt{\frac{36864}{a^4}-12}$.  

Spectral dimension with an extended probe is given by
\begin{equation}
 D_s = n-\frac{a^2}{8\sigma}-\frac{a^2 c^4}{96 k^4}\sigma-\frac{a^2 n}{\sigma}
\end{equation}
which has an extra term, $-\frac{a^2 n}{\sigma}$ compared to eqn.(\ref{sd4}). Note that this additional term depends on the topological dimension $n$ and it is proportional to $\sigma^{-1}$. This will not change the general feature  of the dimensional flow. 

\section{Conclusion}
In this paper, we have constructed four different modified diffusion equations in the $\kappa$-space-time and using their solutions, analyzed the dimensional flow in the $\kappa$-space-time. In these studies, we have used probes which are point-like as well as probe with finite extension. For all these cases, we get the correct commutative limit, where the spectral dimension match with the topological dimension. In all the four cases studied, the spectral dimension changes with the probe scale. We note that for the three cases studied, (see eqns.(\ref{spedim}, \ref{sd3}, \ref{sd4})), in the high energy limit where $\sigma \rightarrow 0$, the spectral dimension become infinitely negative ($-\infty$). This feature was also observed in\cite{sdkappa}. Thus for these three cases spectral dimension looses its meaning at high energies. By demanding that the spectral dimension should be positive definite we obtain bounds on the deformation parameter in terms of diffusion time in these three cases. In the case of spectral dimension obtained in the eqn.(\ref{sdsq}), we note that as $\sigma \rightarrow 0$ spectral dimension becomes equal to topological dimension. In all the four cases, we see a novel feature of spectral dimension of non-commutative space-time in comparison with the result obtained in\cite{sdkappa} as well as in\cite{amelino, nc1, nc2, nc3}. The new fact emerged here is that the spectral dimension goes to $- \infty$ at low energies (i.e.,$\sigma \rightarrow \infty$). We want to emphasis that this feature is absent in the commutative limit and in the commutative limit we get back the equality between the spectral dimension and the topological dimension at low energies. From eqns.(\ref{spedim}),(\ref{sd3}) and (\ref{sd4}), we see that the spectral dimension increases from $-\infty$ as $\sigma$ rises from zero, reaches a maximum value and then decreases to $-\infty$. The maximum value of spectral dimension in all the three cases is less than the topological dimension. A similar behaviour of spectral dimension, but in a completely different context was reported in\cite{relat}. Here the spectral dimension of commutative space-time has been calculated using Relativistic Schr\"odinger Equation Analytically Continued (RSEAC) and the result is compared with the one derived using Telegraph's Equation (TE). The analysis of \cite{relat} shows that only TE produces the spectral dimension that agrees with the topological dimension in the low energies while both these approaches show a reduction of spectral dimension to two at high energies.

We note that the major difference in the present analysis from the earlier ones is the use of a modified diffusion equation(s). In our case, we have not just used Beltrami-Laplace operator in the usual diffusion equation given in eqn.(\ref{de}), but derived the modified diffusion equation in the $\kappa$-deformed space-time. This is done by applying the Wick's rotation to the $\kappa$-deformed Schr\"odinger equation, obtained by taking the non-relativistic limit of well studied $\kappa$-deformed Klein-Gordon equation. This approach, explicitly introduces finite mass for the particle undertaking diffusion on the deformed space-time. We see from the spectral dimension obtained in eqns.(\ref{spedim}, \ref{sdsq}, \ref{sd3}) and eqn.(\ref{sd4}) that in the limit of a massless probe, the spectral dimension and topological dimension coincides at low energies. We note here that the probes used in earlier studies\cite{amelino, nc1, nc2, nc3, sdkappa} were massless ones. The spectral dimension obtained in eqn.(\ref{sdsq}) shows the interesting property that in the limit of probe mass set to zero, there are no correction to spectral dimension due to the non-commutativity. This feature is unique as the spectral dimension calculated for other three cases, do have $a$ dependent term, even in the limit of vanishing probe mass. 

The diffusion equation constructed and analyzed in section 2.3 (see eqn.(\ref{deqnsq})) do not have any higher derivative terms unlike the other three cases studied here (see eqns.(\ref{diff eqn}), (\ref{eqndiff}) and (\ref{boxdiff})). The deformed diffusion equation given in eqn.(\ref{deqnsq}) is obtained from a specific choice of Laplacian (equivalently Klein-Gordon operator in the $\kappa$-deformed space-time). The fact that in the massless limit of the probe, the spectral dimension is exactly same as the topological dimension for all probe scales show that the non-commutativity between time and space coordinate do not affect the spectral dimension of space-part of $\kappa$-space-time at all.

The eqn.(\ref{diff eqn}) and eqn.(\ref{deqnsq}) are derived from the Wick's rotated non-relativistic limit of two different choices of the $\kappa$-deformed Klein-Gordon equation. In the non-relativistic limit, one neglects higher time derivative terms and thus keeps only higher space derivative terms (if any) appear in the deformed diffusion equation. Thus we do not have any higher time derivatives (equivalently, higher derivatives with respect to $\sigma$) in these two equations. Further, for the specific choice of deformed Klein-Gordon equation used in section 2.3, there are no higher order spatial derivatives (upto second order in $a$). This is why the spectral dimension obtained in eqn.(\ref{sdsq}) has a completely different behaviour at high energies. For both the choice of Beltrami-Laplace operator considered in section 3, higher derivatives with respect to spatial as well as Euclidean time coordinates are present and they do appear in the corresponding diffusion equations (see eqn.(\ref{eqndiff})) and (\ref{boxdiff}).  

It is interesting to note that the three diffusion equations leading to negative spectral dimension of high energies all have the higher derivative terms. It has been known that such equations result in negative return probabilities \cite{gc}. In our formulation, $\kappa$-deformed diffusion equations are written down in the commutative space-time. All the effects of non-locality inherent in the non-commutative space-time are contained in the $a$-dependent terms of the deformed diffusion equation. As it is clear, these terms are all higher order derivatives and thus non-local (except for the case studied in section 2.3. As discussed above, the higher time derivative drops out in the non-relativistic limit and this explain why non-commutativity do not play any role in the limit of vanishing mass of the probe for the spectral dimension obtained in eqn.(\ref{sdsq}).). The $\kappa$-deformed Laplacian we used do have higher derivative terms. These terms summarise the non-local effects of the non-commutativity of the space-time. In the momentum space representation of Laplacian, this non-locality appear as higher power terms of momentum\cite{amelino, nc1, nc3}. Laplacians with higher derivatives were also analyzed in\cite{gc, gc2, mz1, mz2}.

The negative value of spectral dimension we see in our analysis might be a reflection of the higher derivative terms (and thus related to the inbuilt non-locality of non-commutative space-time). But the higher derivative terms in the Laplacian (equivalently, Beltrami-Laplace operator) is a characteristic feature of $\kappa$-deformed space-time. Here we have taken a perturbative approach in the analysis of spectral dimension. A detailed analysis of the issue of higher derivatives require a field theoretic re-interpretation going beyond the usual diffusion equation\cite{gc, gc2}. The issues related to higher derivative terms and that of the negative return probability have been analysed in\cite{gc} and, field theoretical re-interpretation of spectral dimension as a possible way to avoid the negative return probability was introduced. For the spectral dimension calculated in eqn.(\ref{spedim}), eqn.(\ref{sd3}) and eqn.(\ref{sd4}), by imposing the requirement that the spectral dimension should be positive definite at high energies translate in to the conditions $a^2 < \frac{72 \sigma}{45+\sigma^2}$, $a^2 < \frac{96 \sigma}{147+\sigma^2}$,$a^2 < \frac{384 \sigma}{12+\sigma^2}$ respectively. This feature suggest the possibility of multiscale structure of the space-time at high energies and such possibilities have been pointed out earlier\cite{gc2}. These issues are being investigated now.

{\noindent{\bf Acknowledgements}}: AV thank UGC, India, for support through BSR scheme.

\end{document}